\documentclass[12pt]{JHEP3}
\usepackage{mathrsfs}
\usepackage{amsmath}
\usepackage{epsfig}
\title{ A Physical Approach to Polya's Conjecture}
\author{Jingbo Wang \\
  Department of Applied Physics,
 Xi'an Jiaotong University,
 Xi'an, 710049, People's Republic of China\\
 \email{ shuijing31@gmail.com }}

\abstract { The similarity between the Polya's conjecture and the
Bonomol'nyi bound remind us to consider a physical approach to
Polya's conjecture. We conjecture a duality between the waves and
the soliton solutions on the surface. We consider the special case
in the disc.}

 \preprint{} \dedicated{} \maketitle
\begin{document}
\section{Introduction}
Let $\Omega  \subset R^2$ be a domain with a boundary $
\partial \Omega$.
Define the Dirichlet Laplacian operator $\Delta$. Then the
eigenvalue problem of the Laplacian on the $\Omega$ is
\begin{equation} \label{eq:1}
\left\{ \begin{aligned}
         \ \Delta \Psi _n  + \lambda _n \Psi _n  = 0
\\
                  \
\Psi _n \left| {(\partial \Omega )} \right. = 0
                          \end{aligned} \right.
                          \end{equation}
 Then Polya's conjecture\cite{P} says that: for every $\lambda_n$, it obey the
 inequality:
 \begin{equation}\label{eq:2}
    \lambda _n  \ge \frac{{4\pi }}{A}n.
 \end{equation}
 Where $A$ is the area of the domain.
Polya himself proved that this conjecture is right for the tiling
domain, but the general case for the arbitrary domain is still open.
Up to now, the best result is obtained by Li and Yau\cite{LY}:
\begin{equation}\label{eq:3}
   \sum\limits_{j = 1}^n {\lambda _j }  \geqslant \frac{{2\pi n^2 }}
{A},n \in \mathbb{N}
\end{equation}

The main goal of this paper is to suggesting a physical approach to
Polya's conjecture, based on the similarity between the inequality
\ref{eq:2} and the Bonomol'nyi bound in soliton theory\cite{H1}.
\section{Nonlinear sigma model}
In this section, we brifely review the nonlinear $O(3)$ $\sigma$
model in $(2+1)$ dimension and its soliton solution\cite{Co}. It
involve three real scalar fields $ \phi (x^\mu  ) \equiv \{ \phi _a
(x^\mu  ),a = 1,2,3,x^\mu   = (t,x,y)\} $ with the constrain:
\begin{equation}\label{eq:4}
    \phi _a \phi _a  = 1.
\end{equation}
Thus, the fields lie on the unit sphere $S_2$.

Subject to this constrain the lagrangian density reads
\begin{equation}\label{eq:5}
    \mathscr{L} = \frac{1} {4}\partial _\mu  \phi _a \partial ^\mu  \phi _a.
\end{equation}
which is invariant under global $O(3)$ rotations in internal space.
From the lagrangian we can get the equation of motion
\begin{equation}
 \partial ^\mu  \partial _\mu  \phi _a  - (\phi _b \partial ^\mu  \partial _\mu  \phi _b )\phi _a  =
 0.
\end{equation}
which for the static case reduces to
\begin{equation}\label{eq:7}
\Delta \phi _a  - (\phi _b \Delta \phi _b )\phi _a  = 0.
\end{equation}

From the constrain equation \ref{eq:4}, we can get
\begin{equation}\label{eq:8}
   \phi _b \Delta \phi _b  =  - \nabla \phi _b  \bullet \nabla \phi
   _b.
\end{equation}
so the equation \ref{eq:7} changes to
\begin{equation}\label{eq:9}
    \Delta \phi _a  + (\nabla \phi _b  \bullet \nabla \phi _b )\phi _a
= 0.
\end{equation}
Define the energy density $ \mathscr{E}(x,y) = \nabla \phi _b
\bullet \nabla \phi _b $, then the potential energy is given by
\begin{equation}\label{eq:10}
  V = \frac{1} {4}\int {(\nabla \phi _b  \bullet \nabla \phi _b }
)dxdy = \frac{1} {4}\int {\mathscr{E}(x,y)} dxdy.
\end{equation}
Notice the similarity between the equation \ref{eq:1} and
\ref{eq:9}.

The problem is completely specified by giving the boundary
condition. We take
\begin{equation}\label{eq:11}
  \mathop {\lim }\limits_{r \to \infty } \overrightarrow \phi
(r,\theta ) = \overrightarrow \phi  ^0.
\end{equation}
where the unit vector $\overrightarrow \phi  ^0$ is a constant
vector. We will take it to be $\overrightarrow \phi  ^0  = (0,0,1)$.

For a soliton solution, we have the Bonomol'nyi bound for the energy
of the soliton,
\begin{equation}\label{eq:12}
  E = V \geqslant 2\pi \left| Q \right|
\end{equation}
where the $Q$ is called the topological charge, or winding number:
\begin{equation}\label{eq:13}
   Q = \frac{1} {{4\pi }}\int_\Omega  {\overrightarrow \phi   \bullet
(\partial _x \overrightarrow \phi  \times \partial _y
\overrightarrow \phi )dxdy} \in \mathbb{N}.
\end{equation}

The mean value of the energy density of the soliton with winding
number $Q=n$ will be
\begin{equation}\label{eq:14}
\overline {\mathscr{E}}  = \frac{{4V}} {A} \geqslant \frac{{8\pi }}
{A}\left| Q \right| = \frac{{8\pi }} {A}n
\end{equation}

If we can find some function between the eigenfunction $\Psi_n$ for
the domain and the $n-$soliton solution $\overrightarrow \phi  ^n $
 of the sigma model on the domain,
 \begin{equation}\label{eq:141}
  \Psi _n  = F(\overrightarrow \phi  ^n ),
\end{equation} that is the duality between the
 waves on the domain and the solitons on the domain, maybe we can
 find the proof of the Polya's conjecture.
\section{The Polya's conjecture on the disc}
In this section, we consider the special case of the Polya's
conjecture, the conjecture on the disc with the radius $R_0$.

The eigenvalue problem of the Laplacian operator can be solved
explicitly. The eigenfunction can be written as
\begin{equation}\label{eq:15}
 \Psi _{mn} (r,\theta ) = cJ_m (\lambda _{mn} r)(a\cos m\theta  +
b\sin m\theta ).
\end{equation}
where $J_m(x)$ is the Bessel function, and the eigenvalues
\begin{equation}\label{eq:16}
   \lambda _{mn}^2  = (\alpha _m^n /R_0)^2
\end{equation}
where $\alpha_m^n$ is the $n-$zero of the $J_m(x)$. The radial
function $R(r)$ satisfy the Bessel function:
\begin{equation}\label{ew:17}
    R'' + \frac{{R'}}{r} + (\lambda _m^2  - \frac{{m^2 }}{{r^2 }})R =
    0.
\end{equation}
This equation also can be obtained by the energy functional
\begin{equation}\label{eq:18}
 E = \int_0^{R_0 } {(R'^2  + \frac{{m^2 R^2 }}{{r^2 }}
 - \lambda _m^2 R^2)
} rdr
\end{equation}
through the Eular-Lagrange equation.

For a special eigenfunction $J_m (\lambda _{mn} r)$, the energy
functional is given by
\begin{equation}\label{eq:19}
E = \int_0^{R_0 } {(\lambda _{mn}^2 (J'_m (\lambda _{mn} r))^2 +
(\frac{{m^2 }}{{r^2 }} - \lambda _{mn}^2 )J_m^2 (\lambda _{mn} r))}
rdr = 1/2\int_0^{\lambda _{mn} R_0 } {(J_{m - 1}^2 (x) + J_{m + 1}^2
(x) - 2J_m^2 (x))xdx}
\end{equation}

Next we consider the nonlinear sigma model on the disc. With the
stereographic projection of the $S_2$, we can get
\begin{equation}\label{eq:20}
   \overrightarrow \phi   = \frac{1}{{1 + uu^* }}(u + u^* , - i(u - u^*
),uu^*  - 1)
\end{equation}
and the complex function $u = \frac{{\phi _1  + i\phi _2 }}{{1 -
\phi _3 }}.$

With the hedgehog ansatz \[ \overrightarrow \phi  (r,\theta ) =
(\sin f(r)\cos n\theta ,\sin f(r)\sin n\theta ,\cos f(r)),
\]where the function $f(r)$ is called the profile function and satisfy some boundary condition, which is \[
f(0) = \pi ,
\] and \[
\mathop {\lim }\limits_{x \to R_0 } f(r) = 0.
\]
 The $n$ is the winding number.
 we get \[ u(r,\theta ) = \tan
(f(r)/2)\exp (in\theta )
\]
and the equation of motion for the complex field is
\begin{equation}\label{eq:21}
 \Delta u - \frac{{2\nabla u \bullet \nabla u}} {{1 + uu^* }}u^*  =
 0.
\end{equation}

So the equation for the $f(r)$ is
\begin{equation}\label{eq:22}
  f'' + \frac{{f'}} {r} - \frac{{n^2 \sin 2f}} {{2r^2 }} = 0.
\end{equation}
Obviously this equation is a nonlinear differential equation and can
be solved numerically (with the shooting method). If the value of
$f(r)$ is obtained then we can use it to calculate the static energy
and so on. As before, this equation can be obtained by the energy
functional
\begin{equation}\label{eq:22}
    E = 2\pi\int_0^{R_0 } {(f'^2 }  + \frac{{n^2 \sin ^2 f}} {{r^2
    }})rdr.
\end{equation}

Notice the similarity and the difference between the two energy
functional \ref{eq:18} and \ref{eq:22}.
\section{The duality}
We want to build the duality between the eigenfunctions and the
soliton solutions on the disc, and from this duality we relate the
eigenvalues of the Laplacian operator to the energy of the dual
soliton, and then from the Bonomol'nyi bound we will get the Polya's
conjecture.

In the disc case, we assume the dual soliton correspondence to the
eigenfunction$ \Psi _{mn}$ is the $N-$ soliton $ \overrightarrow
\phi _N$, then the natural question is how to relate those two
functions and get the function $N(m,n)$. Since we have no explicit
form of the soliton solution, we must guess. Finally we want to
build the relation between the energy of those two solutions, such
as
\begin{equation}\label{eq:23}
 E_{wave}  \geqslant E_{soliton}  \geqslant 2\pi N.
\end{equation}
and we need much more work to achieve this goal.

\section{Conclusion}
In this paper, we just give a general framework for solving the
Polya's conjecture. We hope that our method can be applied to other
related problems in the geometrical analysis, such as: the first
eigenvalue obey the inequality \[ \lambda _1  \geqslant \frac{a}
{{r_\Omega ^2 }},
\]
and the gap between the first two nonzero eigenvalues. We notice
that the eigenvalues have the dimension of the energy density.

On the physical side, some people believe that the fundamental
particles in our universe should be described by the soliton
solution of some nonlinear equation\cite{F}, such as the Skyrme
model for the hadrons in QCD\cite{S}. The duality between the wave
and the soliton can shine some light on those ideas.

For the Neumann boundary, the the Polya's conjecture has the form of
\[
\mu _n  \leqslant \frac{{4\pi }} {A}n.
\]
We didn't find any similar physical results. On the other hand, if
we can find the dual Dirichlet system such that they obey \[ \lambda
_n + \mu _n  = \frac{{8\pi }} {A}n,
\]
then from the Polya's conjecture on the Dirichlet boundary we can
get the problem for the Neumann boundary. The T-duality in string
change the Dirichlet boundary to the Neumann boundary, so we guess
this duality should play some role in the Neumann boundary problem.
\acknowledgments{This work was partly done at Beijing Normal
University. The author will thank the the financial support from the
KITPC in Beijng, China.}
\bibliography{polya}
\end{document}